\pdfoutput=1
\documentclass{PoS}
\usepackage[utf8]{inputenc}
\usepackage{microtype}
\usepackage{amsmath}
\usepackage{booktabs}
\usepackage{braket}
\usepackage{siunitx}
\usepackage{pgf}
\usepackage[caption=false]{subfig}

\newcommand{\Order}[1]{\mathcal{O}\left(#1\right)}
\newcommand{\dd}{\mathrm{d}}
\newcommand{\lalg}[1]{\mathfrak{#1}}
\newcommand{\con}{\text{con}}
\DeclareMathOperator{\tr}{tr}

\newcommand{\dmu}{\hat{\mu}}
\newcommand{\dnu}{\hat{\nu}}

\title{Testing the Witten--Veneziano mechanism\\with the Yang--Mills gradient flow on the lattice}

\ShortTitle{Testing the Witten--Veneziano mechanism with gradient flow on the lattice}

\author{\speaker{Marco Cè}\\
        Scuola Normale Superiore, Pisa \& INFN - Sezione di Pisa\\
        E-mail: \email{marco.ce@sns.it}}

\author{Cristian Consonni, Georg P. Engel and Leonardo Giusti\\
        Università di Milano-Bicocca \& INFN - Sezione di Milano\\
        E-mail: \email{cristian.consonni@gmail.com}, \email{georg.engel@mib.infn.it}, \email{leonardo.giusti@mib.infn.it}}

\abstract{We present a precise computation of the topological charge distribution in the $SU(3)$ Yang--Mills theory. It is carried out on the lattice with high statistics Monte Carlo simulations by employing the clover discretization of the field strength tensor combined with the Yang--Mills gradient flow. The flow equations are integrated numerically by a fourth-order structure-preserving Runge--Kutta method. We have performed simulations at four lattice spacings and several lattice sizes to remove with confidence the systematic errors in the second (topological susceptibility $\chi_t^\text{YM}$) and the fourth cumulant of the distribution. In the continuum we obtain the preliminary results $t_0^2\chi_t^\text{YM}=\num{6.53(8)e-4}$ and the ratio between the fourth and the second cumulant $R=\num{0.233(45)}$. Our results disfavour the $\theta$-behaviour of the vacuum energy predicted by dilute instanton models, while they are compatible with the expectation from the large-$N_c$ expansion.}

\FullConference{The 32nd International Symposium on Lattice Field Theory,\\
		23-28 June, 2014\\
		Columbia University New York, NY}

\begin{document}

\section{Introduction}
The Witten--Veneziano mechanism~\cite{Witten1979a,Veneziano1979} is a solution of the $U(1)_A$ problem, which originates from the experimental observation that in the QCD spectrum there are eight light pseudoscalar mesons, while the ninth, the flavour singlet $\eta'$, is much heavier and cannot be interpreted as a pseudo Nambu--Goldstone boson of chiral symmetry breaking. Indeed, the flavour singlet chiral symmetry $U(1)_A$ is broken by an anomaly. The Witten-Veneziano mechanism is based on the large-$N_c$ expansion~\cite{Hooft1974}, where $N_c$ is the number of colours. In the limit of $N_c\to\infty$ the chiral anomaly vanishes, so $U(1)_A$ symmetry is restored and the $\eta'$ becomes a pseudo Nambu-Goldstone boson, exactly massless in the chiral limit.

The topological susceptibility $\chi_t$ is the two-point function, at zero momentum, of the topological charge density $q(x)$
\begin{equation}
  \chi_t = \int \dd^4x\, \braket{q(x)q(0)}, \qquad q(x) = \frac{1}{64\pi^2} \epsilon_{\mu\nu\rho\sigma} F_{\mu\nu}^a(x) F_{\rho\sigma}^a(x).
\end{equation}
The Witten--Veneziano mass formula relates the anomalous contribution to the $\eta'$ mass to the topological susceptibility in Yang--Mills theory
\begin{equation}
\label{eq:eta_mass_formula}
  M_{\eta'}^2 = \frac{2N_f}{F_{\pi}^2} \chi_t^\text{YM} + \Order{\frac{1}{N_c^2}}.
\end{equation}
A different explanation for the the $U(1)_A$ problem was also proposed by 't Hooft~\cite{Hooft1976}, based on the dilute instanton gas model, a semiclassical approximation of the topological charge distribution of Yang--Mills theory. This approximation fails in the infrared regime of the theory, thus it is not expected to be an accurate description of the mechanism that gives mass to the $\eta'$ meson.

The second moment of topological charge distribution, i.e., the topological susceptibility, is not predicted by the dilute instanton gas model. However, the predictions of Witten--Veneziano and 't Hooft mechanisms differ for higher moments of the distribution. We define $R$ as the ratio between the fourth and the second cumulant of topological charge $Q$ distribution
\begin{equation}
\label{eq:R_def}
  R = \frac{\braket{Q^4}^\con}{\braket{Q^2}}, \qquad \braket{Q^4}^\con = \braket{Q^4} - 3\braket{Q^2}^2.
\end{equation}
Large-$N_c$ arguments predict the ratio $R$ to be of order $1/N_c^2$, while 't Hooft dilute instanton gas model predicts $R=1$.

In this talk, we present preliminary results of the topological charge distribution using high statistics Monte Carlo simulations of $SU(3)$ Yang--Mills theory on the lattice. Our main result is the ratio $R$ in Eq.~\eqref{eq:R_def}, but as a byproduct of the computation we also estimate the second moment of the distribution, the topological susceptibility, since it is related by the Witten--Veneziano mechanism to the mass of the $\eta'$ boson.

\section{Lattice regularization}
We use the discretization of Yang--Mills theory given by the Wilson plaquette action. Our observables are built from the symmetrized \emph{clover} definition of the field strength tensor $F_{\mu\nu}^a$ on the lattice
\begin{equation}
\label{eq:field_strength_tensor_clover}
  F_{\mu\nu}^a(x) = \frac{1}{8a^2} \mathcal{P}^a\left( Q_{\mu\nu}(x) - Q^\dagger_{\mu\nu}(x) \right),
\end{equation}
where $\mathcal{P}^a(X)=-2\tr XT^a$ projects a $N\times N$ matrix to $\lalg{su}(N)$ and $Q_{\mu\nu}(x)$ is the \emph{clover} term
\begin{equation}
  \begin{split}
    Q_{\mu\nu}(x) = &\, U_\mu(x)                 U_\nu(x + a\dmu)                 U^\dagger_\mu(x + a\dnu)       U^\dagger_\nu(x)  \\
                  + &\, U_\nu(x)                 U^\dagger_\mu(x - a\dmu + a\dnu) U^\dagger_\nu(x - a\dmu)       U_\mu(x - a\dmu)  \\
                  + &\, U^\dagger_\mu(x - a\dmu) U^\dagger_\nu(x - a\dmu - a\dnu) U_\mu(x - a\dmu - a\dnu)       U_\nu(x - a\dnu)  \\
                  + &\, U^\dagger_\nu(x - a\dnu) U_\mu(x - a\dnu)                 U_\mu(x + a\dmu - a\dnu)       U^\dagger_\mu(x). \\
  \end{split}
\end{equation}
The observables we measure are the lattice gauge field energy density
\begin{equation}
  E = \frac{1}{4V} \sum_x F_{\mu\nu}^a(x) F_{\mu\nu}^a(x)
\end{equation}
and the topological charge
\begin{equation}
\label{eq:topological_charge}
  Q = \frac{1}{64\pi^2} \sum_x \epsilon_{\mu\nu\rho\sigma} F_{\mu\nu}^a(x) F_{\rho\sigma}^a(x).
\end{equation}
As it is, the two-point function built from this definition of $Q$ is problematic since the sum over coincident points includes divergent contact terms. In this work, we use the definition~\eqref{eq:topological_charge} of $Q$ combined with the Yang--Mills gradient flow. At positive flow-time, it is not affected by this problem and has a well defined continuum limit~\cite{Luescher2011}. However, it is still not proven that this definition satisfies the singlet axial Ward Identity as required by the Witten--Veneziano mechanism~\cite{Giusti2002}.

\subsection{The Yang--Mills gradient flow}
The Yang--Mills gradient flow~\cite{Luescher2010,Luescher2010a} is the solution to the initial value problem
\begin{equation}
\label{eq:gradient_flow_compact}
  \dot{V}(x,t) = -g^2 \{ \partial_{x,\mu} S_\text{W}[V(t)] \} V(x,t), \qquad V(x,0) = U(x),
\end{equation}
where $V(x,t)$ is the gauge field depending on spacetime $x$ and a fifth coordinate, the \emph{flow-time} $t$, and the link differential operator $\partial_{x,\mu}=T^a\partial_{x,\mu}^a$ is defined in~\cite{Luescher2010a}. Towards larger flow-time, the action decreases monotonically and the gauge field is smoothed within a radius of $\sqrt{8t}$. The topological charge $Q(t)$ is defined applying the bosonic definition~\eqref{eq:topological_charge} to positive flow-time gauge field configurations $V(t)$.

\subsection{Fourth-order Runge--Kutta method}

\begin{figure}[t]
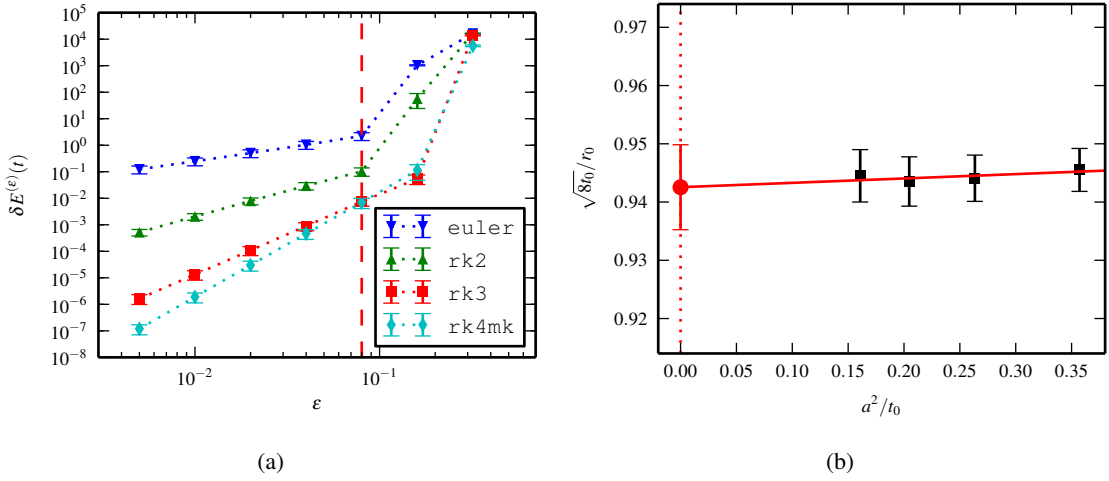

  \subfloat[\label{fig:RK_comparison}]{\input{graphics/RK_comparison/RK_comparison.pgf}}
  \subfloat[\label{fig:t0_scale}]{\input{graphics/t0_scale/t0_scale.pgf}}
  \caption{(a) Comparison of the numerical integration methods. The systematic error $\delta E^{(\epsilon)}(t)$ is estimated from the difference between $E^{(\epsilon)}(t)$, evolved with step size $\epsilon$, and $E^{(\epsilon/2)}(t)$, evolved with step size $\epsilon/2$. (b) Continuum limit of $\sqrt{8t_0}/r_0$.}
  \label{fig:RK_and_t0}
\end{figure}

To perform the numerical integration of the gradient flow, we implement a fourth-order \emph{structure-preserving} Runge--Kutta--Munthe-Kaas (RKMK) method~\cite{Munthe-Kaas1998,Munthe-Kaas1999}. Such a numerical integrator is an algorithm that preserves exactly the geometric structure of the differential equation to integrate, which in this case is the $SU(3)$ Lie group structure of links of the gauge field configuration. The method of choice is
\begin{equation}
\label{eq:RKMK_gradient_flow}
  \begin{aligned}
    W_1              &= V(t), \\
    W_2              &= \exp\left\{ \frac{1}{2}Z_1 \right\} V(t), \\
    W_3              &= \exp\left\{ \frac{1}{2}Z_2 + \frac{1}{8}[Z_1, Z_2] \right\} V(t), \\
    W_4              &= \exp\left\{ Z_3 \right\} V(t), \\
    V(t+a^2\epsilon) &= \exp\left\{ \frac{1}{6}Z_1 + \frac{1}{3}Z_2 + \frac{1}{3}Z_3 + \frac{1}{6}Z_4 - \frac{1}{12}[Z_1, Z_4] \right\} V(t),
  \end{aligned}
\end{equation}
where
\begin{equation}
  Z_i = \epsilon Z\left[W_i\right] = -\epsilon g^2 \{ \partial_{x,\mu} S_\text{W}[W_i] \}, \qquad Z_i \in \lalg{su}(3)
\end{equation}
and $\epsilon$ is the integration step size. The evolution step~\eqref{eq:RKMK_gradient_flow} is applied to every link in the lattice. Its implementation requires space in memory for an auxiliary gauge field and three $\lalg{su}(3)$ Lie algebra fields. In Figure~\ref{fig:RK_comparison} the RKMK method is compared to lower-order Runge--Kutta methods, such as the third order method found in~\cite{Luescher2010a}. $\lalg{su}(3)$ Lie algebra matrix exponentials are computed exploiting the Cayley--Hamilton theorem.

\section{Simulations and results}
We simulate two series of lattices, with details in Table~\ref{tab:simulation_details_and_results}. The lattices in the $A_1,\dotsc,F_1$ series are at fixed lattice spacing of about \SI{0.1}{\femto\meter} in physical units, with increasing volume from $(\SI{1.0}{\femto\meter})^4$ to $(\SI{1.6}{\femto\meter})^4$. The lattices in the $B_1,\dotsc,B_4$ series are at fixed physical volume of $(\SI{1.2}{\femto\meter})^4$, with decreasing lattice spacing.

For each lattice, $N_\text{conf}$ gauge field configurations are generated using the Cabibbo--Marinari heatbath update algorithm, combined with some overrelaxation sweeps. $N_\text{it}$ iteration of heatbath plus overrelaxation sweeps of the full lattice are applied between each measurement. We check the autocorrelation of observables measured at flow-time $t_0$ and the topological charge has the greatest autocorrelation time. Approaching the continuum limit, $N_\text{it}$ is increased in order to separate consecutive measurements with at least three times the integrated autocorrelation time of $Q$.

\begin{table}[t]
  \caption{Details of the lattice simulated and bare numerical results measured at $t_0$.}
  \vskip 8pt
  \label{tab:simulation_details_and_results}
  \begin{tabular}{cS[table-format=1.2]S[table-format=1.2(1)]S[table-format=2]S[table-format=1.1]S[table-format=7]S[table-format=3]S[table-format=1.5(2)]S[table-format=1.3(1)]S[table-format=1.2(1)]}
    \toprule
          & {$\beta$} & {$r_0/a$} & {$L/a$} & {$L$[\si{\femto\meter}]} & {$N_\text{conf}$} & {$N_\text{it}$} & {$t_0/a^2$} & {$\braket{Q^2(t_0)}$} &  {$R(t_0)$}  \\
    \midrule
    $A_1$ &    5.96   &       5.01(2)     &        10       &         1.0     &       36000       &       30   &      2.995(4)       &    0.701(6)   & 0.39(3) \\
    $B_1$ &    5.96   &       5.01(2)     &        12       &         1.2     &      144000       &       30   &      2.7984(9)      &    1.617(6)   & 0.19(2) \\
    $C_1$ &    5.96   &       5.01(2)     &        13       &         1.3     &      280000       &       30   &      2.7908(5)      &    2.244(6)   & 0.18(2) \\
    $D_1$ &    5.96   &       5.01(2)     &        14       &         1.4     &      505000       &       30   &      2.7889(3)      &    3.028(6)   & 0.21(2) \\
    $E_1$ &    5.96   &       5.01(2)     &        15       &         1.5     &      880000       &       30   &      2.7889(2)      &    3.982(6)   & 0.20(2) \\
    $F_1$ &    5.96   &       5.01(2)     &        16       &         1.6     &     1500000       &       30   &      2.78867(16)    &    5.167(6)   & 0.16(2) \\
    \midrule
    $B_2$ &    6.05   &       5.84(2)     &        14       &         1.2     &      144000       &       60   &      3.7960(12)     &    1.699(7)   & 0.24(3) \\
    $B_3$ &    6.13   &       6.63(3)     &        16       &         1.2     &      144000       &       90   &      4.8855(15)     &    1.750(7)   & 0.22(3) \\
    $B_4$ &    6.21   &       7.47(4)     &        18       &         1.2     &      144000       &      250   &      6.2191(20)     &    1.741(7)   & 0.20(3) \\
    \bottomrule
  \end{tabular}
\end{table}

The step size of the integration method is constant for the complete flow on each configuration, but changes between them. For each configuration, the systematic error on the observables $Q^2$ and $Q^4$ is estimated, and the numerical integration is repeated with an halved step size if error bounds are not met. These bounds are chosen to have an upper bound for the systematic error on $R$ which is one order of magnitude smaller than the target statistical error.

\subsection{Reference flow-time}

The reference flow-time $t_0$ is defined through the implicit equation~\cite{Luescher2010a}
\begin{equation}
\label{eq:t0_def}
  \left. t^2 \Braket{E(t)} \right|_{t=t_0} = 0.3,
\end{equation}
with the value $0.3$ chosen such that the smoothing range $\sqrt{8t_0}$ is of the order of the fundamental low-energy scale of the theory. For each lattice we compute the value of $t_0$ from the evolution of the observable $\braket{E(t)}$ with the flow-time $t$. The observables are measured with a flow-time resolution of $0.08a^2$, thus a linear interpolation of data points is used to extrapolate the value of $t$ which solve Eq.~\eqref{eq:t0_def}. Our results are given in Table~\ref{tab:simulation_details_and_results}. Figure~\ref{fig:t0_scale} shows the continuum limit of the dimensionless quantity $\sqrt{8t_0}/r_0$, with Sommer scale $r_0$ values given in~\cite{Guagnelli1998}. The $a=0$ value is
\begin{equation}
  \frac{\sqrt{8t_0}}{r_0} = \num{0.943(7)}, \qquad \frac{t_0}{r_0^2} = \num{0.1110(17)}. %0.9425456+-0.0072935   0.1110490+-0.0017186
\end{equation}
Using the value $r_0F_K=\num{0.4146(94)}$~\cite{Garden2000} and the experimental input $F_K=\SI{160(2)}{\mega\electronvolt}$, we have $r_0=\SI{0.511(13)}{\femto\meter}$, thus the $t_0$ scale in physical units is
\begin{equation}
\label{eq:t0_phys}
  t_0 = \left(\SI{0.170(5)}{\femto\meter}\right)^2. %0.1703935+-0.0046043
\end{equation}
The observables values in lattice units reported in Table~\ref{tab:simulation_details_and_results} are extrapolated to the value of $t_0$ measured on the same lattice using a linear interpolation.

\subsection{Thermodynamic limit}

\begin{figure}[t]
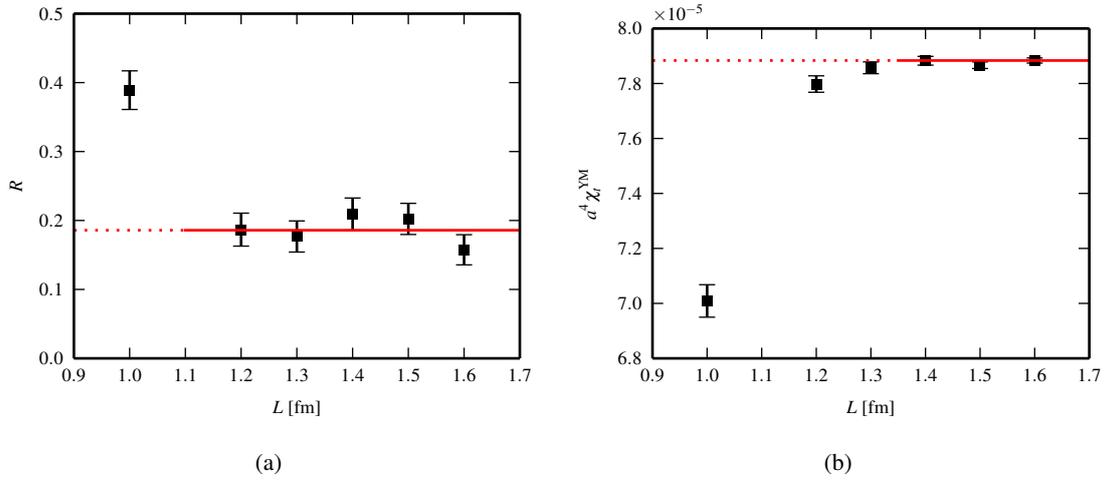

  \subfloat[\label{fig:thermodynamic_limit_R}]{\input{graphics/volume_effects_R.pgf}}
  \subfloat[\label{fig:thermodynamic_limit_chi}]{\input{graphics/volume_effects_chi.pgf}}
  \caption{(a) Thermodynamic limit of $R$. (b) Thermodynamic limit of $a^4\chi_t^\text{YM}$.}
  \label{fig:thermodynamic_limit}
\end{figure}

In Figure~\ref{fig:thermodynamic_limit_R} we show the finite volume effects on $R$ from simulations at fixed lattice spacing. Starting from a volume of $(\SI{1.2}{\femto\meter})^4$, finite volume effects are compatible with the statistical error. Since for $R$ the latter scales with $V$, a number of configurations increasing with $V^2$ is necessary to maintain the same statistical error on $R$. Therefore, to limit the computational cost of the simulations we decided to study the continuum limit at a fixed physical volume of $(\SI{1.2}{\femto\meter})^4$. From Figure~\ref{fig:thermodynamic_limit_chi}, we see that, since the statistical error on the observable in lattice units $a^4\chi_t^\text{YM}$ is smaller, finite volume effects are visible at $V=(\SI{1.2}{\femto\meter})^4$, thus a correction is needed to have a result for the topological susceptibility valid in the thermodynamic limit.

\subsection{Continuum limit}

\begin{figure}[t]
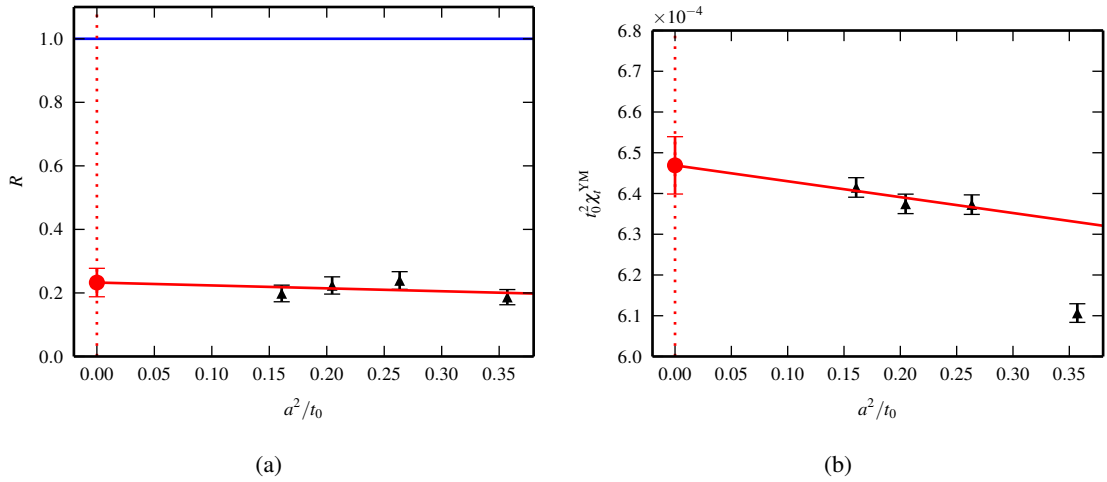

  \subfloat[\label{fig:continuum_limit_R}]{\input{graphics/continuum_limit_R.pgf}}
  \subfloat[\label{fig:continuum_limit_chi}]{\input{graphics/continuum_limit_chi.pgf}}
  \caption{(a) Continuum limit of $R$. In blue the dilute instanton gas model prediction $R=1$. (b) Continuum limit of $t_0^2\chi_t^\text{YM}$.}
  \label{fig:continuum_limit}
\end{figure}

Figure~\ref{fig:continuum_limit_R} shows the continuum limit of $R$. The result, valid in the thermodynamic and continuum limit, is
\begin{equation}
\label{eq:continuum_limit_R}
  R = \num{0.233(45)}. %0.23270 +- 0.04476
\end{equation}
Figure~\ref{fig:continuum_limit_chi} shows the continuum limit of the dimensionless quantity $t_0^2\chi_t^\text{YM}$. Excluding the value of the coarser lattice from the fit, the value at $a=0$ is $t_0^2\chi_t^\text{YM}=\num{6.47(7)e-4}$, which needs to be corrected for the finite volume effects in Figure~\ref{fig:thermodynamic_limit_chi}. The corrected value, valid in the thermodynamic and continuum limit, is
\begin{equation}
\label{eq:continuum_limit_chi}
  t_0^2\chi_t^\text{YM} = \num{6.53(8)e-4}, \qquad \chi_t^\text{YM} = \left( \SI{185(5)}{\MeV} \right)^4, %(6.528895+-0.075719)e-4
\end{equation}
using $t_0$ in physical units given by Eq.~\eqref{eq:t0_phys}.

\section{Conclusions}

We studied the topological charge distribution with an unprecedented precision and our result~\eqref{eq:continuum_limit_R} is the the first result for the ratio $R$ with systematic and statistical errors under control. The value~\eqref{eq:continuum_limit_R} agrees with the result of~\cite{Giusti2007}, it is inconsistent with the dilute instanton gas model prediction of $R=1$, and it is compatible with the large-$N_c$ prediction of being of order $1/N_c^2$. We also measured the topological susceptibility with unprecedented precision. The value~\eqref{eq:continuum_limit_chi} is compatible with other lattice results~\cite{DelDebbio2005,Luescher2010b}.

\bibliographystyle{JHEP}
\bibliography{biblio.bib}

\end{document}